\documentclass[iop,apjl]{emulateapj}
\usepackage{epsfig}
\usepackage{amsmath, amsthm, amssymb}
\usepackage[plainpages=false, colorlinks=true, anchorcolor=blue, linkcolor=blue, citecolor=blue, bookmarks=false, urlcolor=blue]
{hyperref}
\usepackage{color}
\usepackage{microtype}
\usepackage{float}
\usepackage{verbatim}
\usepackage{wrapfig}
\usepackage{graphicx}
\usepackage[caption = false]{subfig}
\usepackage{multirow}
\usepackage{hhline}
\usepackage{booktabs}
\usepackage{array}
\usepackage{bm}

\newcommand{\Msun}{$M_\sun$}

\newcommand{\shortauth}{S. Swihart}
\newcommand{\slugcom}{Draft version - Accepted to ApJ}
\slugcomment{\slugcom}

\lefthead{\sc \footnotesize \slugcom \hfill \shortauth}
\righthead{\sc \footnotesize \slugcom \hfill \shortauth}

\begin{document}

\title{A New Likely Redback Millisecond Pulsar Binary with a\\ Massive Neutron Star: 4FGL J2333.1--5527}

\author{Samuel J. Swihart\altaffilmark{1},
Jay Strader\altaffilmark{1},
Ryan Urquhart\altaffilmark{1},
Jerome A. Orosz\altaffilmark{2},
Laura Shishkovsky\altaffilmark{1},
Laura Chomiuk\altaffilmark{1},
Ricardo Salinas\altaffilmark{3},
Elias Aydi\altaffilmark{1},
Kristen C. Dage\altaffilmark{1},
Adam M. Kawash\altaffilmark{1}}

\affil{ 
  \altaffilmark{1}{Center for Data Intensive and Time Domain Astronomy, Department of Physics and Astronomy,\\Michigan State University, East Lansing, MI 48824, USA}\\
  \altaffilmark{2}{Department of Astronomy, San Diego State University, 5500 Campanile Drive, San Diego, CA, 92182, USA}\\
  \altaffilmark{3}{NSF's National Optical-Infrared Astronomy Research Laboratory/Gemini Observatory, Casilla 603, La Serena, Chile}\\
}

\begin{abstract}
We present the discovery of a likely new redback millisecond pulsar binary associated with the \emph{Fermi} $\gamma$-ray source 4FGL J2333.1--5527. Using optical photometric and spectroscopic observations from the SOAR telescope, we identify a low-mass, main sequence-like companion in a 6.9-hr, highly inclined  orbit around a suspected massive neutron star primary. Archival \emph{XMM-Newton} X-ray observations show this system has a hard power-law spectrum $\Gamma = 1.6\pm0.3$ and $L_X \sim 5 \times 10^{31}$ erg s$^{-1}$, consistent with redback millisecond pulsar binaries. Our data suggest that for secondary masses typical of redbacks, the mass of the neutron star is likely well in excess of $\sim1.4\,M_{\odot}$, but future timing of the radio pulsar is necessary to bolster this tentative conclusion. This work shows that a bevy of nearby compact binaries still await discovery, and that unusually massive neutron stars continue to be common in redbacks.
\end{abstract}

%Our data suggest that the system harbors a massive ($1.88\pm 0.24\,M_{\odot}$) neutron star whose properties can be much better constrained with its future detection as a millisecond radio pulsar.

\section{Introduction}
Prior to the launch of the \emph{Fermi} Gamma-ray Space Telescope \citep{Atwood09} in 2008, most known millisecond pulsar (MSP) binaries had white dwarf companions in relatively wide orbits ($P_{\textrm{orb}}\gtrsim$ 5 d). Systems like these represent the end point of the MSP recycling process  \citep{Tauris06}, in which a stellar companion accretes matter and angular momentum onto the neutron star surface, spinning it up to rapid spin periods.

However, recent multiwavelength follow-up observations of newly-discovered $\gamma$-ray sources from the \emph{Fermi} Large Area Telescope (LAT) have revealed a substantial population of MSPs with non-degenerate companions. These systems generally display radio eclipses due to material from the secondary, making pulsations difficult to detect in blind radio timing searches. They are usually classified by the mass of their secondaries: black widows have very lightweight ($M_c \lesssim$ 0.05 \Msun), semi-degenerate companions that are highly ablated by the pulsar, while redbacks have more massive ($M_c \gtrsim$ 0.1 \Msun) main sequence-like companions \citep{Roberts11,Strader19} and typically show more extensive radio eclipses \citep[e.g., ][]{Camilo15, Cromartie16, Keane18}. The redback class has drawn considerable attention in recent years due to three systems that have been observed to transition between an accretion-powered disk state and a rotationally-powered pulsar state on short ($\lesssim$\,month) timescales, proving the suspected evolutionary link between low-mass X-ray binaries and MSPs \citep{Archibald09, Papitto13, Bassa14} and showing that the MSP recycling process in redbacks may be ongoing.

Concerted multiwavelength observations of binary MSPs have enabled mass measurements of a number of neutron stars, placing tight constraints on the equation of state of dense matter and improving our understanding of fundamental physics at supranuclear densities \citep{Steiner13,Ozel16}. 
Using 32 precise neutron star mass measurements, \citet{Antoniadis16} found that the MSP mass distribution is strongly bimodal, with a narrow peak near the canonical $\sim$1.4\;\Msun~and a second, broad peak around $\sim$1.8\;\Msun. Intriguingly, the neutron star masses in the approximately two dozen known redbacks are consistent with being drawn almost exclusively from the more massive second peak of this proposed neutron star mass distribution \citep{Strader19}.

For MSP binaries with non-degenerate companions, the neutron star mass estimates are typically dependent on the binary inclinations inferred from fitting the orbital variations in the optical light curves. When the companion is  significantly heated by the pulsar (true for all black widows and many redbacks), this modeling is complex, and hence the 
inclinations can be plagued by substantial systematic uncertainties. This leads, in turn, to conflicting mass estimates \citep[e.g.,][]{Schroeder14,Romani15,Sanchez17,Linares18}. However, in systems that are nearly edge-on, valuable constraints on the neutron star mass are possible that are practically independent of the less certain light curve modeling.

In this paper, we present the discovery of the compact binary associated with the \emph{Fermi} $\gamma$-ray source 4FGL J2333.1--5527, showing that it is likely a redback MSP binary with a low-mass main-sequence companion in a highly inclined orbit around a massive neutron star. 

\section{Observations}

\subsection{The $\gamma$-ray Source \& Optical Discovery}
\label{sec:gamma}
4FGL J2333.1--5527 is a bright unassociated \emph{Fermi}-LAT $\gamma$-ray source with an overall detection significance of 19.4$\sigma$ in the 0.1--100 GeV energy range based on eight years of survey data \citep{4FGLcatalog}. The source is persistent in $\gamma$-rays and has appeared in the previous 1FGL \citep{Abdo10}, 2FGL \citep{Nolan12}, and 3FGL \citep{Acero15} catalogs. The source has a significantly curved $\gamma$-ray spectrum with no evidence for variability, consistent with most other $\gamma$-ray MSPs \citep{Abdo13}.

\begin{figure*}[t!]
\begin{center}
  \subfloat{\includegraphics[width=0.5\textwidth]{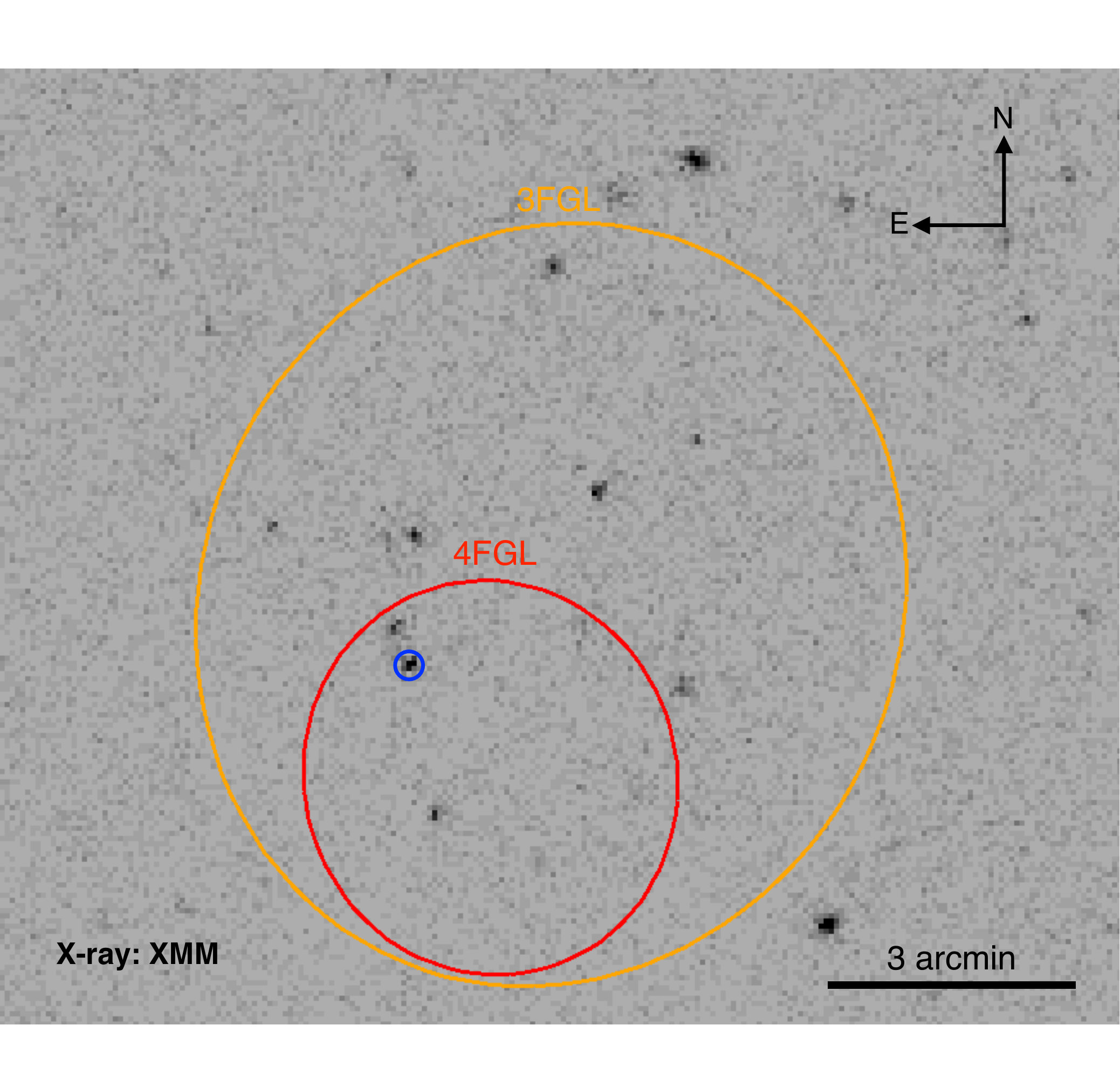}}
  \subfloat{\includegraphics[width=0.5\textwidth]{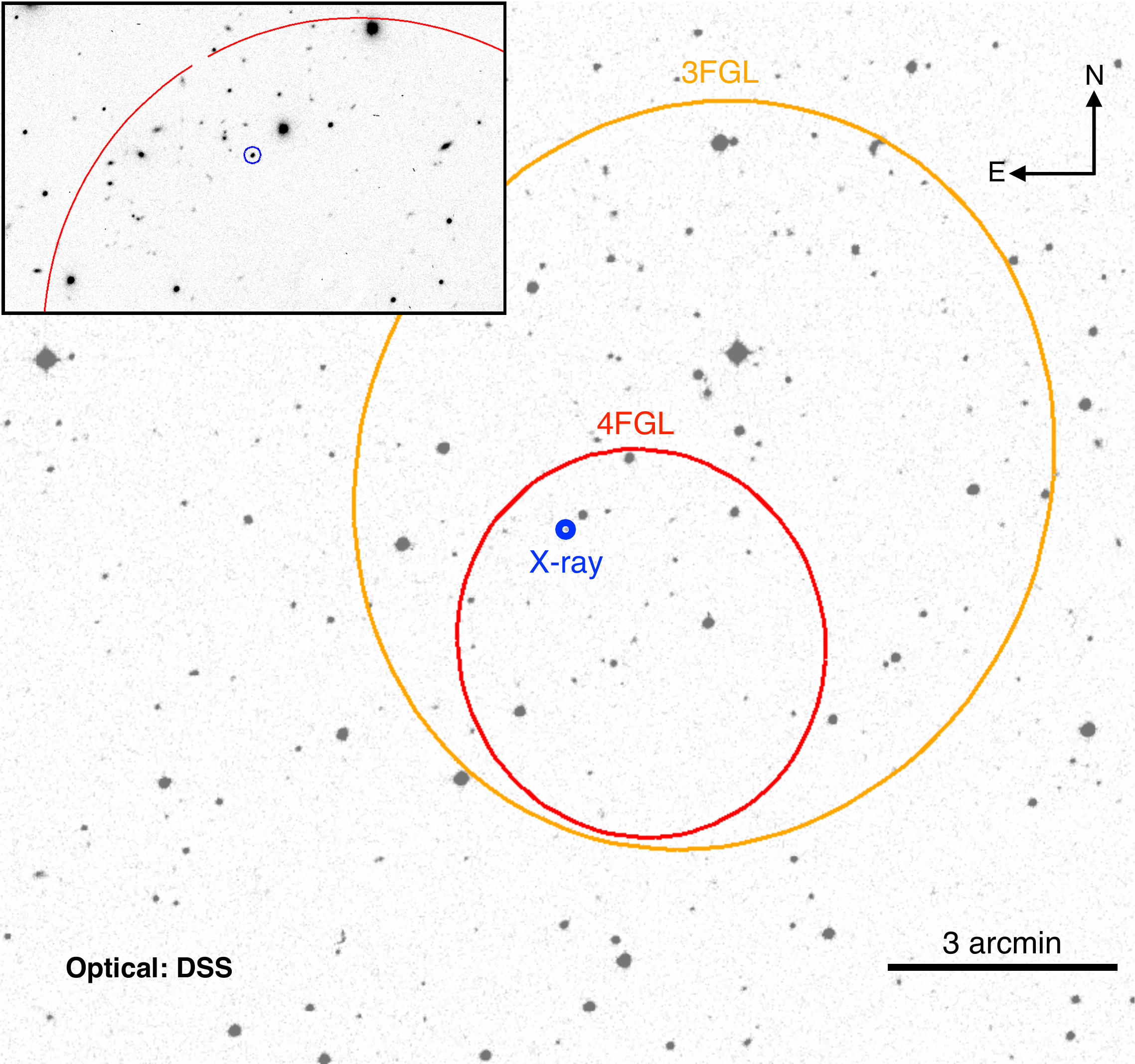}}
\caption{Left: \emph{XMM-Newton}/EPIC X-ray image of the field of 4FGL J2333.1--5527 with the 95\% error regions from the 4FGL (red) and 3FGL (orange) catalogs overlaid. The subject of this work is marked with the blue circle. Right: Optical DSS image of the field with the location of the X-ray source marked in blue. The inset displays a zoomed-in view of one of our SOAR $i'$ images showing the variable optical counterpart associated with the X-ray source.}
\label{fig:finders}
%\vspace{4mm}
\end{center}
\end{figure*}

The 4FGL 95\% positional error ellipse lies entirely within the 3FGL region and is $\sim$70\% smaller in area (Figure~\ref{fig:finders}). The 0.1--100 GeV flux corresponds to a luminosity of $4.4 \times 10^{33}\,(d/3.1 \;\rm{kpc})$ erg s$^{-1}$. This distance is derived from the photometric light curve fitting in \S~\ref{sec:ELCmodel} and is a bit more distant than the median redback distance (1.8 kpc; \citealt{Strader19}). In any case, none of our central results depend on the exact distance, and given the faintness of the secondary, it is very unlikely this distance is off by more than a factor of $\sim 2$.

The subject of this paper is the strong candidate optical counterpart to 4FGL J2333.1--5527. We first identified this source as a possible variable star within the 4FGL error ellipse due to its unusually large \emph{Gaia} $G$ mag uncertainty compared to nearby sources of similar brightness. This source is listed in \emph{Gaia} DR2\footnote{https://www.cosmos.esa.int/web/gaia/dr2} with a brightness $G \sim 20.3$ mag and is located at a J2000 sexagesimal position of (R.A., Dec.) = (23:33:15.967, --55:26:21.06). For the remainder of this work, we refer to the optical and X-ray source (see below) as J2333.

\subsection{X-ray Observations}
\label{sec:xrayobs}

Prior to the launch of \emph{Fermi}, the field containing 4FGL J2333.1--5527 was observed briefly and serendipitously with \emph{XMM-Newton} on 2007 Nov 29 (ObsID 0505381001; PI Boehringer). We downloaded the publicly-available European Photon Imaging Camera (EPIC) observation from NASA's High Energy Astrophysics Science Archive Research Centre (HEASARC) archive\footnote{https://heasarc.gsfc.nasa.gov/docs/archive.html}. Data were reprocessed using standard tasks in the Science Analysis System ({\small SAS}) version 18.0.0 software package. We used standard flagging criteria \verb|FLAG=0|, in addition to \verb|#XMMEA_EP| and \verb|#XMMEA_EM| for pn and MOS, respectively. Patterns 0-4 were used for pn and 0-12 for MOS. We found no evidence of strong, extended background flares, giving a total exposure time of $\sim$8.7 ksec.

The optical variable discussed above matches the position of the brightest X-ray source within the error ellipse of 4FGL J2333.1--5527 (Figure~\ref{fig:finders}), and this is the source we analyze. We extracted background-subtracted spectra and light curves using circular source extraction regions of radius 15\arcsec~and local background regions three times larger. For our timing analysis, barycentric corrections were applied to all event files using the {\small SAS} task \texttt{barycen}. Individual MOS1, MOS2 and pn light curves were extracted using the {\small SAS} tasks \texttt{evselect} and \texttt{epclccorr} before being combined into a single EPIC light curve using the FTOOLS \citep{Blackburn95} task \texttt{lcmath}. Similarly, for our spectral analysis, individual MOS1, MOS2 and pn spectra were extracted using standard \texttt{xmmselect} tasks and then combined using \texttt{epicspeccombine} in order to create a weighted-average EPIC spectrum. Spectral fitting was performed using XSPEC \citep{Arnaud96} version  12.10.1 (\S~\ref{sec:Xrayspec}). Due to the limited number of source counts, XSPEC's implementation of Cash statistics \citep{Cash79} modified for a background-subtracted spectrum, W-statistics, was used to test the goodness of our spectral fits.

\subsection{Optical Observations}
\label{sec:optobs}

\subsubsection{SOAR Photometry}
\label{sec:photometry}
We observed J2333 on 2018 Aug 27, Sep 19, and Oct 23 and 2019 Aug 19 and Sep 15 using SOAR/Goodman in imaging mode. On each of our 2018 nights we took a series of 180 sec exposures with the SDSS $i'$ filter \citep{Fukugita96}; during the 2019 nights we also included alternating exposures in $g'$.
Each image covered a circular 7.2\arcmin~diameter field and was binned 2$\times$2 giving a final plate scale of 0.3\arcsec/pixel. Typical seeing was 0.8\arcsec, 1.4\arcsec, and 1.0\arcsec~on the 2018 August, September, and October nights, respectively, and 1.1\arcsec~and 1.0\arcsec~on the 2019 August and September nights.

The raw images were corrected for bias and then flat-fielded with a combination of dome and sky flats using standard packages in \texttt{IRAF} \citep{Tody86}. We performed aperture photometry to obtain the instrumental magnitudes of the target, which were then calibrated using the $g'$- and $i'$-band magnitudes of a number of nearby comparison stars taken from The Blanco Cosmology Survey \citep{Desai12}. To ensure that any variability we observed from our target was real, we examined the light curves of our comparison stars to choose only the most stable stars, finding 14 reference stars in $g'$ and 24 in $i'$. Images that were affected by clouds or that were taken under very poor seeing conditions ($\gtrsim$1.8\arcsec) were removed. The final photometric sample consists of 121 and 272 measurements in $g'$ and $i'$, respectively. The mean observed magnitudes of the target source (not corrected for extinction) are $g' = 21.123$ mag and $i' = 19.868$ mag, with median per epoch errors of $\sim$0.039 mag and $\sim$0.013 mag in $g'$ and $i'$, respectively.

For the remainder of the paper, we refer to the bright phases of the optical light curve as the ``dayside'', corresponding to when we observe the inner heated face of the tidally locked companion, while the ``nightside'' corresponds the companion's non-illuminated backside (see \S~\ref{sec:LCdescrip}).

\subsubsection{SOAR Spectroscopy}
\label{sec:spectroscopy}
We obtained spectra of J2333 using the Goodman spectrograph \citep{Clemens04} on the SOAR telescope on Cerro Pachon, Chile from 2018 Aug 20 to 2019 Aug 6. All data used a 400 l mm$^{-1}$ grating with an approximate wavelength range of either $\sim 4800$ to 8800 \AA~or a setup about 1000 \AA~bluer. Depending on the seeing, we used slit widths of 0.95\arcsec~or 1.2\arcsec~yielding resolutions of 5.7 or 7.3 \AA~FWHM, respectively. Owing to the faintness of the star, the exposure times were relatively long, 25 or 30 min per spectrum.

The spectra were reduced and optimally extracted in the normal manner. Radial velocities were measured through cross-correlation with bright templates of similar spectral type around the Mg$b$ region of the spectra. 42 spectra had high enough signal-to-noise for velocities to be measurable. The mid-exposure observation times are reported as Modified Barycentric Julian Dates (BJD - 2400000.5 d) on the TDB system \citep{Eastman10}. 

\begin{figure}
	\includegraphics[width=\linewidth]{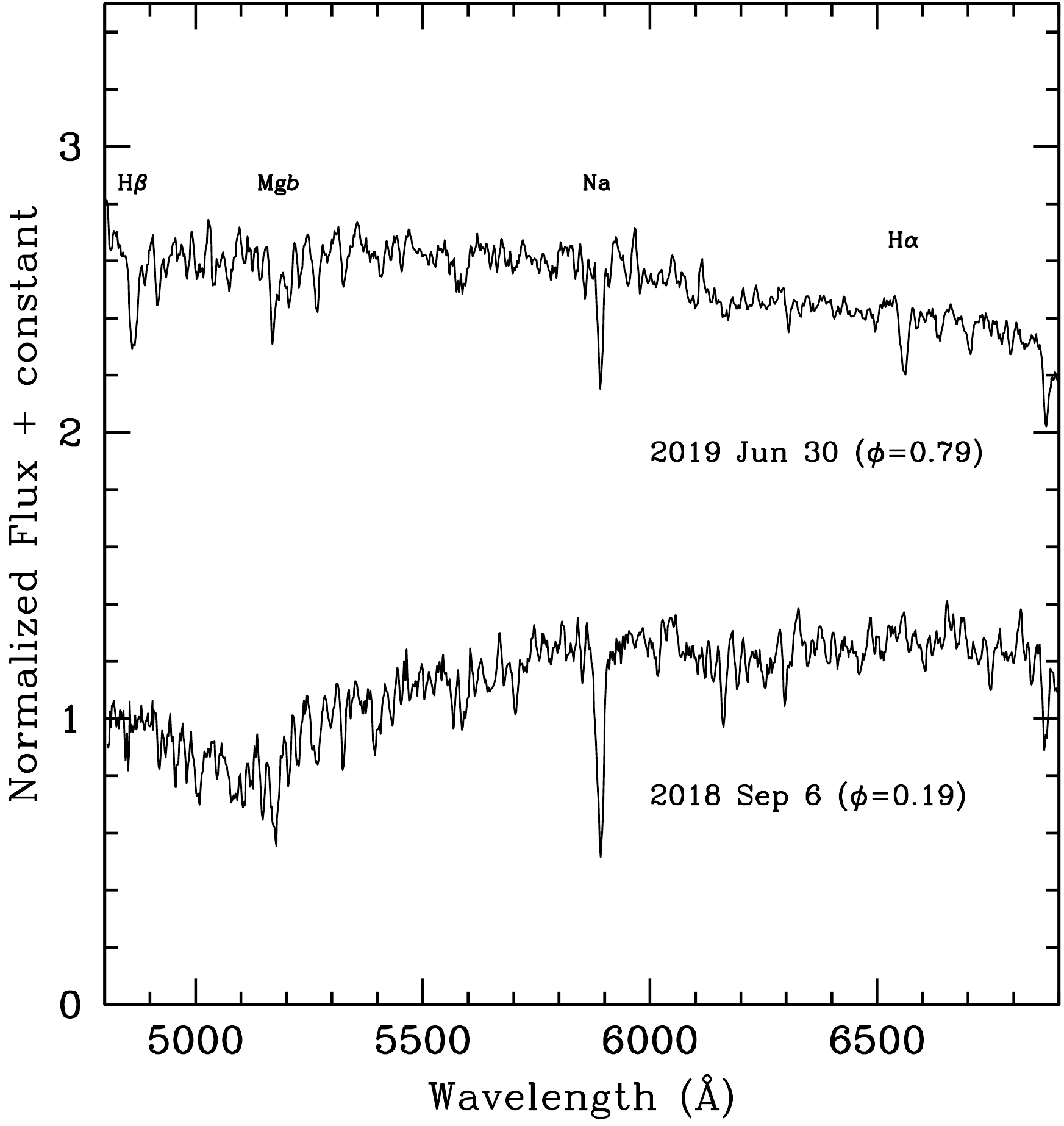}
    \caption{Two of our SOAR spectra showing examples of the varying effective temperature of the companion. The dayside spectrum (top) is consistent with an early-G star, while the nightside spectrum (bottom) is markedly cooler, matching that of a mid-K dwarf. Our other low-resolution spectra typically lie somewhere between these extremes.}
    \label{fig:opt_spectra}
\end{figure}

The nightside spectra are consistent with having spectral types as low as mid-K, while most of the dayside spectra are best matched as early-G stars (Figure~\ref{fig:opt_spectra}). One or two spectra appear even warmer, into the range of an early-F star with weak metal lines, perhaps reflecting a transient episode of increased heating. We show below that the mean colors of the star on its night and day sides are fully consistent with the typical spectra observed.

Nearly all of the spectra cover both H$\alpha$ and H$\beta$, and while there is Balmer absorption present in most of the warmer spectra, there is no clear evidence for emission in any of the spectra. This is unlike some other redbacks, where the Balmer emission is attributed to a wind or shock \citep{Swihart18}.

In some redbacks that show strong evidence for irradiation, the radial velocities depends on the absorption lines used to measure them. For example, Balmer lines will tend to trace the brighter (warmer) portions of the companion, which are closer to the binary center of mass and therefore move slower, whereas metallic lines track the dark (cooler) side, which is moving faster \citep[e.g., ][]{Linares18}. For the few spectra that showed clear H$\alpha$ absorption, we compared the radial velocities of these lines to the Mg$b$ velocities, but did not find evidence for a clear offset between these velocities at any phase.

%We find that the radial velocities depend systematically on the atmospheric absorption lines used to measure them. Namely, the semi-amplitude of the radial velocity curve of J2215 measured with magnesium triplet lines is systematically higher than that measured with hydrogen Balmer lines, by 10%. We interpret this as a consequence of strong irradiation, whereby metallic lines dominate the dark side of the companion (which moves faster) and Balmer lines trace its bright (slower) side.

\section{Results \& Analysis}

\subsection{Modeling the Radial Velocities}
\label{sec:specresults}

\begin{figure}[t]
\includegraphics[width=1.0\linewidth]{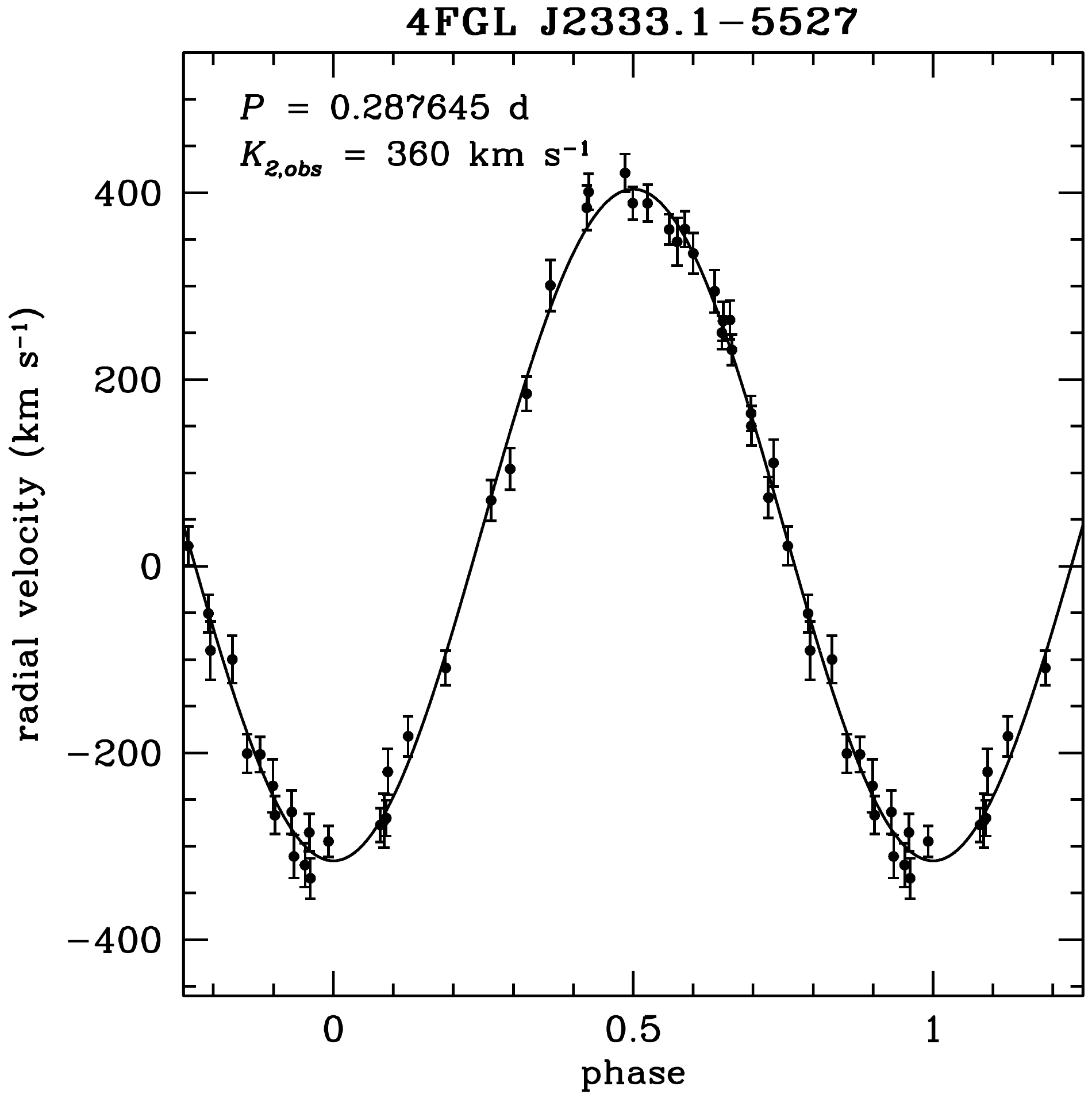}
\caption{Circular Keplerian fit to the observed radial velocities. As discussed in the text, the fit to the ``corrected" velocities has a lower semi-amplitude but is otherwise extremely similar.}
\label{fig:rvmodel}
\end{figure}

We fit Keplerian models to the radial velocities using the Monte Carlo sampler \emph{TheJoker} \citep{Price17}. Initially we fit a four-parameter circular model, finding period $P =  0.2876450(14)$ d, epoch of ascending node $T_0 = 58463.46881(65)$ d, semi-amplitude $K_2 = 360\pm5$ km s$^{\rm{-1}}$, and $v_{sys} = 44\pm4$ km s$^{\rm{-1}}$. These parameters represent a good fit, with $\chi^2 = 36.4$ for 38 d.o.f.~and an rms of 20.5 km s$^{-1}$. This model is plotted with the original velocity data in Figure~\ref{fig:rvmodel} (here and throughout this paper, we use the orbital phase convention where $\phi=0.25$ is the inferior conjunction of the companion).

However, this model of the data may not be complete: the substantial heating of the secondary implied by the light curves (see \S~\ref{sec:LCdescrip} and Figure~\ref{fig:opt_LC}) suggests that heating could also be affecting the measured radial velocities of the secondary.  Such heating can displace the center of light outward from the center of mass, inflating the measured velocity semi-amplitude or causing a false eccentricity \citep{Davey92}. There is no statistical support for an eccentric model: again fitting with \emph{TheJoker}, the posterior for the eccentricity is peaked at 0, with a median posterior value of $e=0.009$. Using median values of this fit reduces the $\chi^2$ value by only 0.2 despite two additional free parameters. As other tests, performing circular fits for relevant subsets of the data (most/least heated phases, or closest to quadrature) also did not substantially affect the measured $K_2$ within its uncertainty. Hence there is little direct evidence from the velocities themselves that they are affected by irradiation.

Nonetheless, to make a first-order estimate of the effect heating might have on our velocities, we use the ELC models calculated to match the light curve in \S~\ref{sec:ELCmodel}. At each phase of the model, we take the intensity-weighted velocity offset of each grid element from the center of mass as viewed by an observer, and combine it with an effective temperature--equivalent width relation for Mg$b$ taken from \citet{Johansson10}. This method is similar to that used by \citet{Shahbaz17} except that we do not set the equivalent widths of the elements heated beyond a certain point to zero, since this is not demanded by the data. Future spectroscopy with a larger telescope would allow improvements in the quality of the spectra that could show the need for additional modeling. 

Table~\ref{tab:RVdata} shows the original velocities as well as those with this correction applied. The strength of the correction is highest near quadrature. Refitting a circular model to the corrected velocities gives parameters identical to the original fit within the uncertainties, excepting the semi-amplitude, which is lower at $K_2 = 338(4)$ km s$^{-1}$. The overall fit is of slightly lower quality ($\chi^2 = 38.2$ for 38 d.o.f.; rms of 20.8 km s$^{-1}$), and indeed there is no particular evidence from the data that this model provides a better fit than the original one. In subsequent sections we note results from both the original and ``corrected" velocities to indicate how the conclusions depend on these effects.

\begin{deluxetable}{crrc}[]
\tablecaption{SOAR Velocities of J2333}
\tablehead{BMJD & RV & RV$_{\textrm{corr}}$ & err \\
        (d) & (km s$^{-1}$) & (km s$^{-1}$) & (km s$^{-1}$)}
\startdata
58345.3394130 & 184.9 & 181.8 & 18.2 \\
58367.2301482 & 401.0 & 386.4 & 19.3 \\
58367.2476498 & 421.3 & 397.9 & 20.1 \\
58367.2763716 & 361.2 & 340.9 & 19.1 \\
58367.2940899 & 250.2 & 238.4 & 18.1 \\
58368.2446582 & --320.0 & --295.9 & 23.5 \\
58368.2942262 & --182.1 & --175.1 & 21.5 \\
58368.3121900 & --108.9 & --106.2 & 18.3 \\
58368.3340083 & 70.6 & 70.0 & 21.9 \\
58372.1703291 & 335.1 & 316.6 & 21.8 \\
58372.1879465 & 263.8 & 253.8 & 20.8 \\
58372.2087764 & 110.7 & 109.0 & 25.1 \\
58372.2263937 & --90.3 & --85.7 & 31.2 \\
58372.3114531 & --220.2 & --208.3 & 24.5 \\
58380.3103683 & --235.3 & --217.0 & 28.4 \\
58380.3278272 & --285.2 & --260.7 & 20.1 \\
58462.0531349 & --277.1 & --263.1 & 18.2 \\
58463.0939823 & 150.3 & 144.7 & 21.3 \\
58463.1114704 & 21.7 & 22.5 & 20.9 \\
58463.1327180 & --99.8 & --90.8 & 25.3 \\
58463.1701621 & --334.3 & --309.7 & 21.4  \\
58484.0827981 & 231.8 & 222.3 & 16.5  \\
58484.1002837 & 73.6 & 71.0 & 22.0 \\
58491.0546756 & --266.7 & --248.0 & 20.2 \\
58491.1070246 & --272.6 & --259.6 & 29.1 \\
58503.0410349 & 347.6 & 325.8 & 25.8 \\
58637.3104413 & 300.8 & 295.2 & 27.5 \\
58637.3279199 & 384.0 & 369.9 & 24.0 \\
58637.3499781 & 388.9 & 364.5 & 17.8 \\
58637.3674571 & 360.7 & 337.6 & 16.2 \\
58637.3893140 & 294.6 & 281.2 & 22.8 \\
58637.4067921 & 163.7 & 158.1 & 18.9 \\
58664.2250896 & --263.3 & --241.1 & 23.3  \\
58664.2426674 & --294.7 & --269.9 & 16.5 \\
58664.2703763 & --269.8 & --257.5 & 19.0 \\
58664.3298441 & 104.2 & 102.2 & 22.2 \\
58664.3956801 & 388.8 & 363.8 & 19.8 \\
58665.2948716 & 262.6 & 251.1 & 21.0 \\
58665.3356255 & --50.6 & --46.3 & 20.2 \\
58665.3605401 & --201.5 & --186.1 & 19.0 \\
58701.3098865 & --200.6 & --188.3 & 20.4 \\
58701.3323067 & --310.8 & --288.2 & 23.1
\enddata
\label{tab:RVdata}
\end{deluxetable}
%\vspace{1cm}

\subsubsection{Minimum Neutron Star Mass}
\label{sec:minmass}

Using the posterior samples from our Keplerian fits to the original radial velocities, we derive the mass function $f(M) = P \, K_2^3 \, (1-e^2)^{3/2}/(2\pi G) = M_{NS} \, \textrm{sin}^3 i/(1+q)^2 = 1.39\pm0.05 \, M_{\odot}$, for mass ratio $q = M_{2}/M_{NS}$ and inclination $i$. For the heating-corrected velocities, the $K_2$ is lower by about 6\%, leading to a mass function of $f(M) = 1.15\pm0.04\,M_{\odot}$. Both values are large compared to typical redbacks, suggesting a relatively edge-on orientation, and in the former case a massive neutron star primary. Assuming the original velocities and the median mass ratio of known redbacks ($q = 0.16$), the minimum mass of the presumed neutron star for an edge-on ($i=90^{\circ}$) orbit is $M_{NS} = 1.87 \pm 0.07\,M_{\odot}$.

The mass ratio is not well-constrained from these data and hence the detection and timing of the neutron star as a pulsar will be necessary to improve our estimate of the primary mass. If we instead assume the minimum possible mass ratio ($q = 0$), the minimum neutron star mass is $M_{NS} = 1.39 \pm 0.05\,M_{\odot}$ for an edge-on orbit, with smaller inclinations resulting in a heavier neutron star. We revisit the mass estimate in the context of the light curve fitting in \S~\ref{sec:ELCmodel}.

\subsection{X-ray Spectrum}
\label{sec:Xrayspec}
\begin{figure}
	\includegraphics[width=\linewidth]{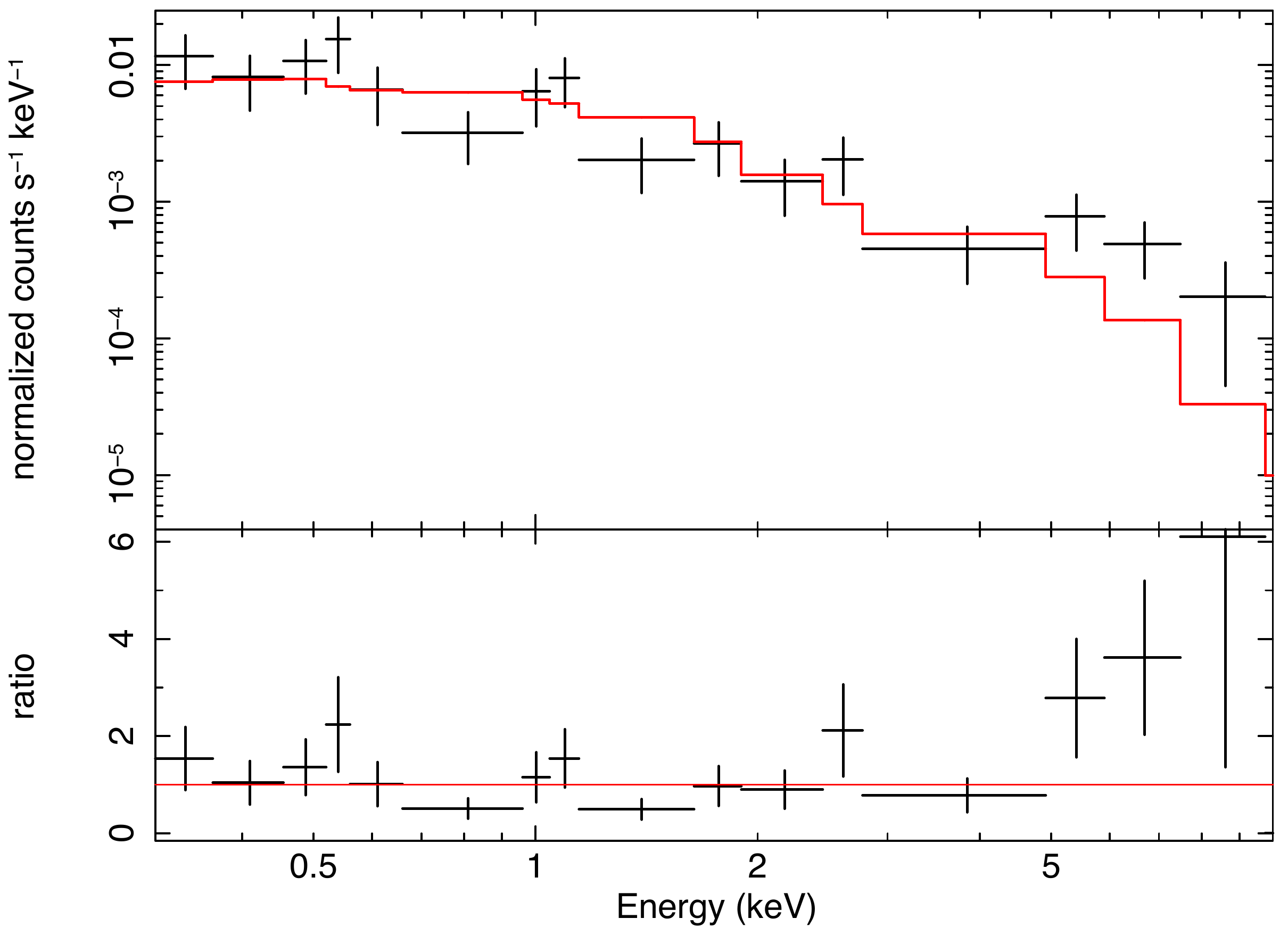}
    \caption{\emph{XMM-Newton}/EPIC MOS+pn spectrum of J2333, with model and model/data ratios. Due to the low number of source counts, the unbinned data were fitted with W-statistics and later rebinned to a minimum significance of $2.2\sigma$ for visualization purposes only. The model (red) shows our best-fit absorbed power-law model (\texttt{TBabs*powerlaw}) with photon index $\Gamma=1.6\pm0.3$.}
    \label{fig:xray_spectra}
\end{figure}

We initially fit the spectrum of J2333 with simple absorbed power-law (\texttt{TBabs*powerlaw}) and blackbody (\texttt{TBabs*bbody}) models. It is immediately apparent that the spectrum is too broad to be fit with a single blackbody (cstat/dof = 171/217), and the absorbed power-law provides an appreciably better fit (cstat/dof = 132/217; Figure \ref{fig:xray_spectra}). Using the power-law model, we find no evidence of additional intrinsic absorption and thus fix the \texttt{TBabs} parameter to the line-of-sight absorption column density $N_{\mathrm H}=1.1\times10^{20}$\,cm$^{-2}$ \citep{Dickey90, Kalberla05, HI4PI16}. The best-fitting photon index is $\Gamma=1.6\pm0.3$, with an unabsorbed $0.3-10$\,keV flux of $\left(4.3^{+1.4}_{-1.1}\right)\times10^{-14}$\,erg s$^{-1}$ cm$^{-2}$. This equates to a luminosity of $L_{\mathrm X}=\left(5.0^{+1.5}_{-1.3}\right)\times10^{31}$\,erg s$^{-1}$, using the reference distance of 3.1 kpc.

We also tried a more complex \texttt{TBabs*(powerlaw + bbody)} model in an attempt to account for the fit residuals found at energies $\gtrsim5$\,keV (Figure \ref{fig:xray_spectra}). A number of binary MSPs have been shown to have evidence for a thermal component in their spectrum \citep[e.g., ][]{Heinke09, Archibald10, Bogdanov11, Roberts14, Roberts15} likely coming from the heated magnetic polar caps on the surface of the neutron star. The introduction of a soft ($kT=0.09\pm0.05$\,keV) blackbody component causes the power-law to become flatter ($\Gamma=1.1\pm0.5$). In this case, we find a slightly higher total unabsorbed $0.3-10$\,keV flux of $\left(5.7^{+2.3}_{-1.7}\right)\times10^{-14}$\,erg s$^{-1}$ cm$^{-2}$, of which the power-law contributes $\approx90\%$. However, this model (cstat/dof = 127/215) is comparable statistically to the pure absorbed power-law model and thus we prefer the simpler model.

\subsection{X-ray Light Curve}
\begin{figure}[t]
\includegraphics[width=1.05\linewidth]{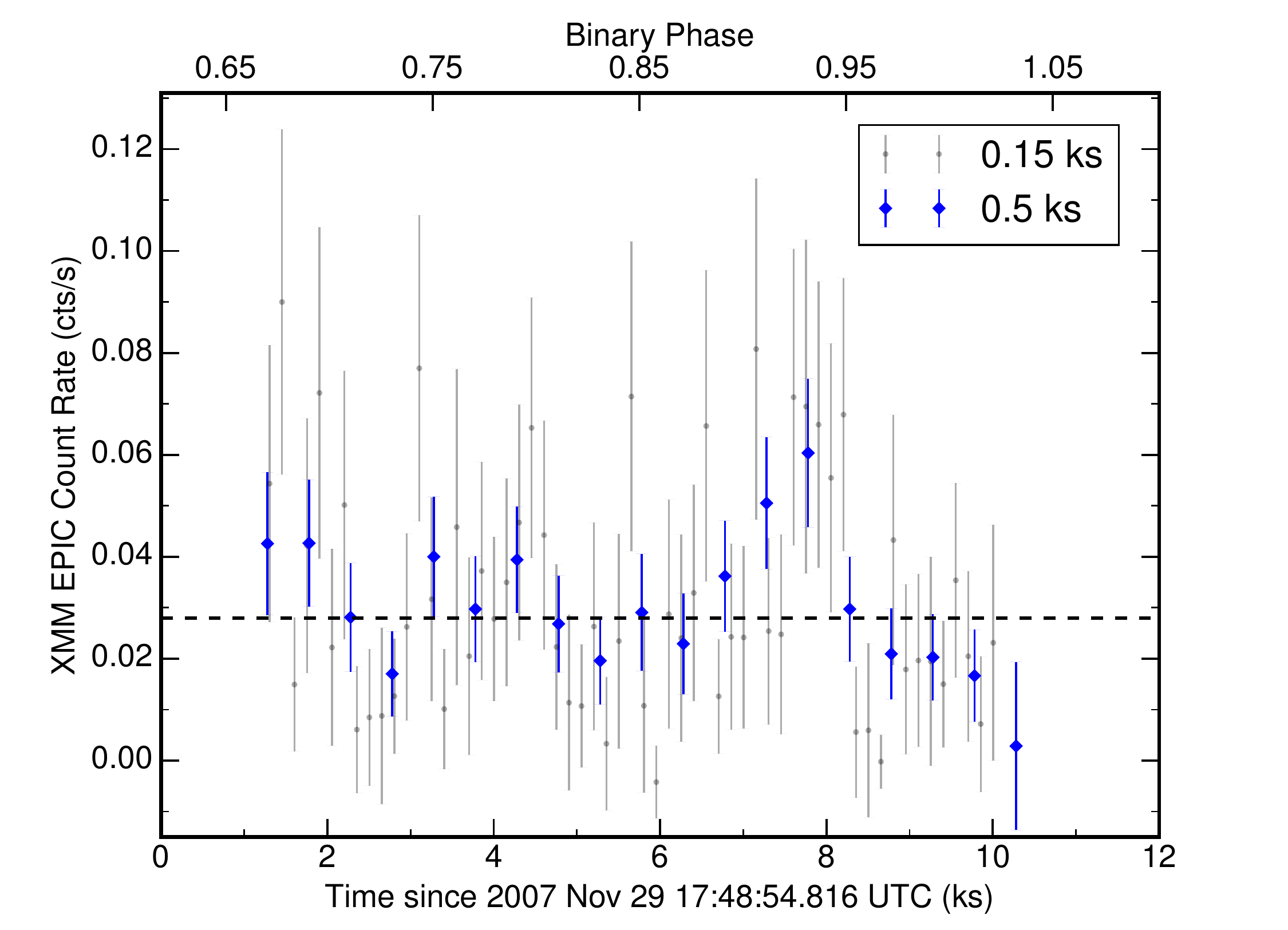}
\caption{The \emph{XMM-Newton}/EPIC light curve between 0.2--10 keV. The observation covers $\sim$40\% of the binary orbit. Barycenter corrections have been applied to the photon arrival times and the data are grouped into 150s (black) and 500s (blue) bins. The weighted average of the 500s binned dataset is shown with a dashed line.}
\label{fig:xmm_LC}
\end{figure}

The background-subtracted X-ray light curve of J2333 is shown in Figure~\ref{fig:xmm_LC}. After applying barycentric corrections, the data were grouped into 150s and 500s time bins. The 8.7 ksec exposure covers approximately 40\% of the orbit from $\phi \sim 0.65-1.05$, and shows considerable variability on short timescales. The black dashed line in Figure~\ref{fig:xmm_LC} represents the weighted average count rate of the 0.5 ks binned dataset and is shown for visualization purposes only.

Although the 0.2--10 keV count rates vary by at least a factor of 3 over this short interval, a $\chi^2$ fit suggests a probability of $\sim$3\% that this data could be produced from a constant flux distribution. Together with the hard X-ray spectrum presented above, this emission can be attributed to an intrabinary shock that occurs at the interface between the wind driven from the companion and the relativistic pulsar wind \citep[e.g.,][]{Gentile14,Romani16,AlNoori18}. A longer observation would allow us to determine if this variability were real and possibly modulated on the orbital period of the binary.

Consistent with the low flux, this source was not detected in a short \emph{Swift/XRT} observation performed as a part of a program monitoring unidentified \emph{Fermi} sources \citep{Stroh13}. This highlights the need for deeper X-ray observations of other unassociated \emph{Fermi} $\gamma$-ray sources.

\subsection{Optical Light Curve}
\label{sec:LCdescrip}
The optical light curves show clear evidence of variability, with $g'$ and $i'$ magnitudes ranging from $\sim$20.4--21.8 and $\sim$19.55--20.35 mag, respectively. Our 2019 Sep data covers slightly more than a full orbital cycle, suggesting the period is approximately 6.9 hours. In order to confirm this estimate, we performed a period search on the full photometric dataset (five epochs) using a Lomb-Scargle periodogram. The highest power peak was found at a period of $P_{\rm{orb}}$ = 0.28764467 d $\approx$ 6.90347 hrs. Since this value is fully consistent with the spectroscopic period, for the remainder of the paper we assume the orbital period and ephemerides (and corresponding uncertainties) derived from the spectroscopy (\S~\ref{sec:specresults}).

\begin{comment}
\begin{figure}[t]
\includegraphics[width=1.05\linewidth]{paperfig_Sep19LC.pdf}
\caption{SOAR $i'$-band light curve of J2333 taken in 2019 Sep, covering just over one full orbital cycle. No data were taken during the gap seen near hour 1.}
\label{fig:timeseriesLC}
\end{figure}
\end{comment}

\begin{figure}[t]
\includegraphics[width=1.05\linewidth]{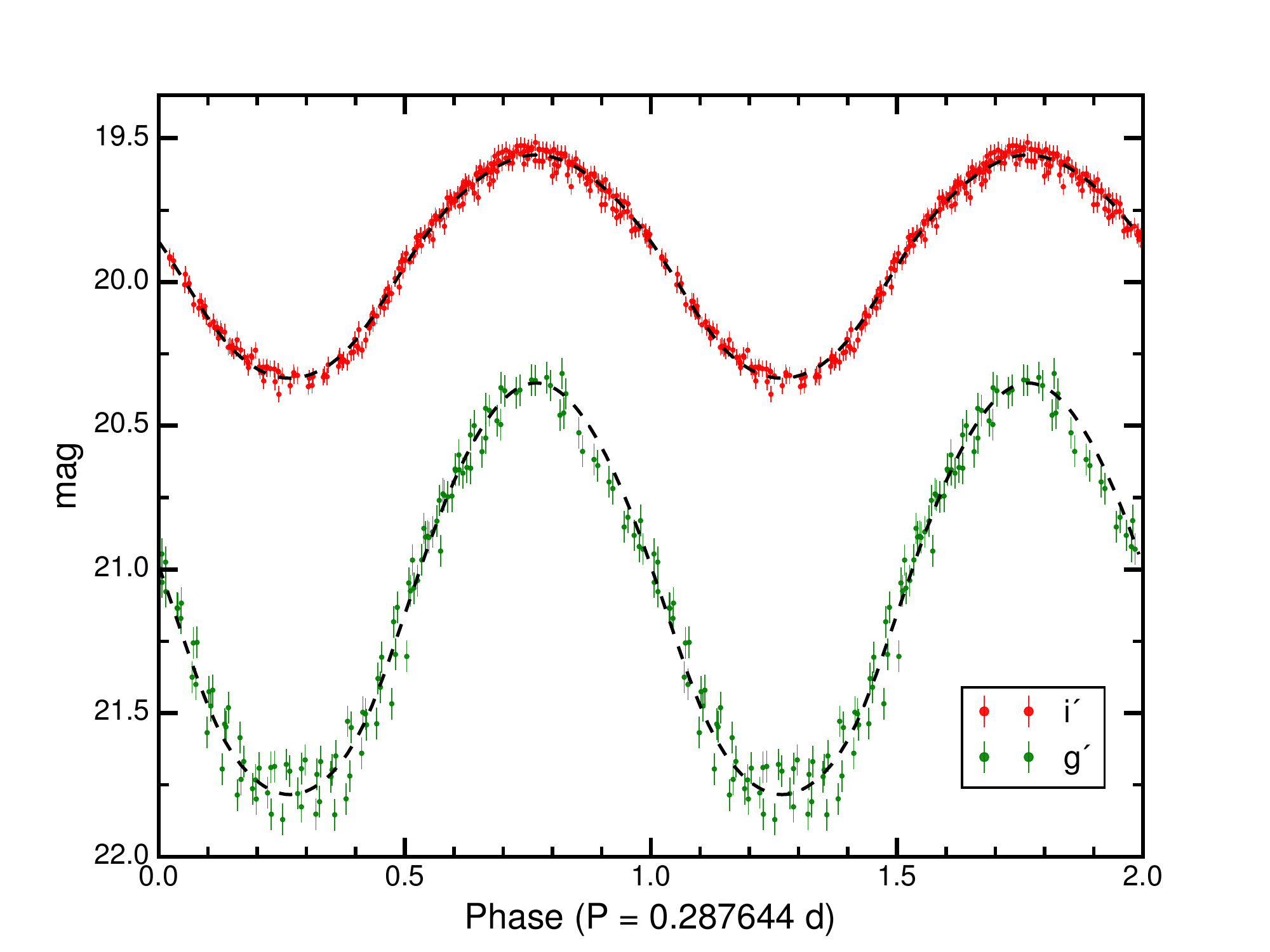}
\caption{SOAR $g'$ and $i'$ photometry of the companion to J2333 along with the best fit \texttt{ELC} model. The data described in \S~\ref{sec:photometry} have been folded on the best-fit period and ephemeris from \S~\ref{sec:specresults}, and the uncertainties have been rescaled as described in the text (\S~\ref{sec:ELCmodel}). Two orbital phases are shown for clarity.}.
\label{fig:opt_LC}
\end{figure}

We calculated the midpoint of each observation and then folded these data on our best fit orbital period and $T_0$. Consistent with our spectroscopic convention, $\phi = 0$ represents the ascending node of the suspected pulsar, such that the secondary lies along the line connecting the primary and Earth at $\phi = 0.25$. The folded light curves are displayed in Figure~\ref{fig:opt_LC}.

Similar to a number of other redback MSPs, the light curves show a broad maximum and narrower minimum per orbital period, consistent with the secondary being heated on its tidally locked ``day'' side by the suspected pulsar primary. This heating is dominating the shape of the light curve over ellipsoidal modulations due to tidal deformation of the secondary, which would manifest in two maxima/minima per orbital cycle.

In redbacks, heating patterns like these typically occur when the pulsar spin-down power is reprocessed in an intrabinary shock formed from the interaction between the pulsar and companion winds \citep[][]{Romani16, Wadiasingh17, Kandel19}. Such a shock would also produce a characteristic hard power-law X-ray spectrum  \citep[e.g.,][]{Roberts15, Werner16}, consistent with our observations (\S~\ref{sec:Xrayspec}).

Overall, the light curves are fairly well-behaved with a few notable exceptions. First, in $i'$ the scatter in the data appears more pronounced after $\phi \sim 0.65$, suggesting increased intrinsic stochastic variability when viewing the inner heated face of the companion. This increased variability also appears in $g'$, so that the observed maximum appears slightly later than the expected $\phi = 0.75$, while the shape and slope of the $g'$ data around $\phi = 0.85-0.95$ suggests the light curve is not well-behaved at these phases.

Lastly, the scatter in the $g'$ data when viewing the companion's nightside ($\phi \sim 0.1-0.4$) is appreciably larger than at any other phases, with the brightness varying by nearly 0.2 mag over timescales of a few minutes. This phenomenology suggests that there is variable heating ongoing on the companion surface similar to what has been observed in a number of other redback MSPs \citep[e.g.,][]{Cho18}. This variability could be explained by a number of physical processes such as magnetic reconnection events between the intrabinary shock and stellar surface, a rapidly varying shock geometry, inhomogeneities, asymmetries, and/or magnetic channeling of the pulsar wind, or migrating star spots on the companion \citep[e.g.,][]{Deneva16, vanStaden16, Halpern17, Sanchez17, Zharikov19}. Together with our observation in \S\ref{sec:spectroscopy} that a few optical spectra have noticeably higher effective temperatures than the others, this variability reinforces the idea that this system undergoes irregular transient heating episodes that affect the tidally locked companion.

\subsubsection{Modeling the Light Curves}
\label{sec:ELCmodel}
Armed with the SOAR $g'$ and $i'$ light curves, we model the system using the Eclipsing Light Curve \citep[\texttt{ELC};][]{Orosz00} code. Throughout our modeling we assume a compact, invisible primary with no accretion disk, and a tidally distorted secondary in a circular orbit. We hold the orbital period $P$ and semi-amplitude $K_2$ to the values found from our best radial velocity curve fits (\S~\ref{sec:specresults}). We also assume the secondary is tidally locked, a valid assumption since the synchronization timescale is $\lesssim 10^4$ yr at the orbital period of J2333 \citep{Zahn77}.

Since we do not yet have an estimate for the mass ratio of the binary, we assume the median value for the known redbacks: $q = M_2/M_{\rm{NS}} = 0.16$ \citep{Strader19}. Leaving the mass ratio as a free parameter results in a wide range of well-fitting models mainly due to the degeneracy between this parameter and the filling factor of the secondary (i.e., a smaller star can accommodate smaller mass ratios and vice versa). Future observations, such as high-resolution spectroscopy to measure the companion's rotational broadening, or a measurement of the projected semi-major axis of the pulsar orbit would allow for a direct determination of the mass ratio and break this degeneracy. 

The overall scale of the system is already set by $P$, $K_2$, and $q$, giving an orbital separation for the secondary of $\sim$2.3 $R_{\odot}$. Finally, the controlling physical parameters of our light curve models are the effective temperature of the companion's nightside $T_2$, the isotropic irradiating luminosity of the suspected pulsar (which we characterize here as the maximum temperature on the day side of the heated secondary) $T_{\rm{day}}$, the Roche lobe filling factor of the companion $f_2$, the binary inclination $i$, and a phase shift $\Delta \phi$. This last parameter can account for any small offsets that might be present between the light and radial velocity curves, which are commonly found in black widow or redback binaries dominated by irradiation \citep[e.g.,][]{Schroeder14,Draghis19, Swihart19}.

As mentioned in \S~\ref{sec:spectroscopy}, we analyzed a number of our low/medium resolution spectra of J2333 that were taken when we were viewing the nightside of the secondary ($\phi = 0.1-0.4$) to constrain $T_{2}$. These spectra are consistent with a $\sim$K5 main sequence star with an effective temperature of $\sim$4400 K.

To confirm this value, we also examined the nightside colors of the companion using our SOAR photometry. After applying extinction corrections using the \citet{Schlafly11} reddening maps, the de-reddened color at $\phi \sim 0.25$ is ($g'-i')_{0} = 1.428$. We compare this value to the theoretical colors of main sequence stars using a 10 Gyr solar metallicity isochrone \citep{Bressan12}, finding $T_2 \sim 4470$K. If we instead assume a more metal-poor star with [Fe/H] = --1, we find a lower $T_2 \sim 4350$K. These values are broadly consistent with our estimate from the spectral type so for the remainder of the analysis, we fix the nightside effective temperature of the secondary to $T_2 = 4400$K.

We initially found models that appeared superficially to provide good fits to the data but had a large formal $\chi^2$/dof = 1702/389, mostly dominated by the small errors on the $i'$ photometry. Furthermore, there are more than twice as many $i'$-band measurements as there are in $g'$, so when modeling the light curves together \texttt{ELC} preferentially finds models that fit the $i'$ data well, while the $g'$ data receives very little weight. In addition, a small fraction of data points have substantially smaller uncertainties than the typical data at that phase, which can dominate the model fitting, essentially forcing the model to go thru (``overfit'') these small error datapoints. This is especially problematic when trying to fit the nightside in $g'$ and the $i'$ data near maximum, where the increased scatter is most pronounced and only a couple datapoints with small errors can dominate the model at these phases.

Therefore, we rescaled the uncertainties so that all intraband errors were equal and adjusted the values so that the total reduced $\chi^2$ of the final model was $\sim 1.0$. This process ensured the photometry in both bands contributed approximately equally to the total $\chi^2$ while also insulating against ``overfitting'' datapoints that had substantially smaller uncertainties than the typical data at that phase. This method has been used in the literature for modeling the light curves of redbacks and achieving fits with reliable uncertainties \citep[e.g.,][]{McConnell15, Linares18, Strader19}.

The best fit \texttt{ELC} model along with the photometry (with rescaled uncertainties) is shown in Figure~\ref{fig:opt_LC}. The best fit model suggests the star is only partially filling its Roche lobe ($f_2 = 0.75 \pm 0.01$), has a dayside temperature of $\sim$5700 K (consistent with our optical spectra; \S~\ref{sec:spectroscopy}), and is in a nearly edge-on orbit ($i \sim 86^{\circ}$).

Using these values together with our spectroscopic results, we infer the mass of the primary associated with a range of reasonable secondary masses. The distribution of redback companion masses can be modeled with a normal distribution: $M_{2} = 0.36 \pm 0.16$ \citep{Strader19}. Utilizing this distribution along with our measurement of the binary mass function $f(M) = 1.39\pm0.05 \, M_{\odot}$ (\S~\ref{sec:minmass}), the primary mass is $M_{NS} = 1.96^{+0.25}_{-0.27}\,M_{\odot}$. In these models the secondary has a radius of $R_2 \sim 0.51\,R_{\odot}$. If instead we use the smaller mass function corresponding to our ``heating-corrected'' radial velocity model, the primary mass is $M_{NS} = 1.70^{+0.23}_{-0.25}\,M_{\odot}$, though as we discuss in \S~\ref{sec:specresults} there is no evidence that this ``heating-corrected'' $K_2$ fits the data better than the original $K_2$ value. Regardless, both these estimates for the primary mass suggest the presumed neutron star is well in excess of the canonical $\sim1.4\,M_{\odot}$.

%assuming normal K2 --> f(M)=1.39 +-0.05
%M2 = 0.36 +- 0.16 --> M_NS = 1.96, 1.69, 2.21 --> 1.96+-0.26
%1.96+0.25-0.27 !!!!!!!!!!!!

%assuming corrected K2 --> f(M)=1.15 +-0.04
%M2 = 0.36 +- 0.16 --> M_NS = 1.70, 1.45, 1.93 --> 1.70+-0.24
%1.70+0.23-0.25 !!!!!!!!!!!!!

%the most likely mass is XXX2 +/- YYY +^ZZZ_1 _ZZZ2, where YYY is the random uncertainty due primarily to the uncertainty in K_2, and ZZZ_1/ZZZ_2 come from using the observed distribution of redback mass ratios (q ~ 0.16 +/- 0.05).

%If we instead assume the minimum known redback mass ratio \citep[$q=0.07$;][]{Strader19} the primary mass is $\sim1.6\,M_{\odot}$.

%assuming normal K2 --> f(M)=1.39
%q = 0.0, M_NS = 1.4
%q = 0.07, M_NS = 1.6
%q = 0.16, M_NS = 1.88
%q = 0.29, M_NS = 2.33

%assuming smaller K2 --> lower f(M)=1.15
%q = 0: M = 1.16
%q = 0.07: M = 1.33
%q = 0.16, M_NS = 1.56
%q = 0.29: M = 1.93

%Using these values together with our spectroscopic results and assuming the rough $q$ distribution for redbacks from \citet{Strader19} ($q=0.16\pm0.07$), the inferred mass of the primary is $M_{NS} = 1.88\pm0.24$ \Msun, consistent with a massive neutron star.

%If we assume the lower value for $K_2$ from \S~\ref{sec:specresults}, the model is largely unchanged except for the component masses, which are instead $M_{NS} = 1.56\pm0.20\,M_{\odot}$ and $M_2 = 0.25\pm0.03\,M_{\odot}.$
%\,M_{\odot}$.

The inclination is not precisely constrained owing to the slight degeneracies between this parameter, the filling factor, and the level of heating. We explored models where $T_{2}$ (and therefore also the relative heating on the dayside) was allowed to vary within the constraints set by our optical spectra, but found that models that both fit the data well and were consistent with our spectra fully agree with our main results within a few percent.
%Models that both fit the data well and are consistent with our optical spectra do not change the final results within a few percent.
%ther than something simple like "we accounted for (uncertainties on) XX and it only changed the final results by YY%”
The best fit value for the inclination and 1$\sigma$ uncertainty is $i = 85\fdg8^{+4.2}_{-19.1}$. Since the currently assumed value is close to edge-on, allowing a wider range of inclinations would allow a more massive neutron star than discussed above. That said, using the original $K_2$ measurement and the median secondary mass for known redbacks, models with $i \lesssim 74^{\circ}$ imply neutron star masses larger than the current Shapiro delay record holder ($2.14^{+0.10}_{-0.09} M_{\odot}$; \citealt{Cromartie19}) and hence are less likely. Lower values of $K_2$ can accommodate a wider range of inclinations.

\subsubsection{Distance}
\label{sec:distance}
Similar to the steps described in \citet{Strader15} and \citet{Swihart19}, we use our best light curve models to infer the distance to the system. We first derive the intrinsic luminosity of the system by assuming $T_2$ and $R_2$ from our best fit model, then apply bolometric corrections in each band as a function of temperature using the same solar metallicity 10 Gyr isochrone listed above. Finally, we compare these predicted absolute magnitudes to the mean de-reddened apparent magnitudes in each band to infer a distance of 3.1 $\pm$ 0.3 kpc. The uncertainty in this estimate includes systematic effects due to the unknown metallicity and Roche lobe filling factor of the star as well as the dispersion between filters. 

A future \emph{Gaia} data release should provide a parallax distance for this source to test the photometric estimate; for the handful of sources with accurate parallax distance measurements, these photometric light curve distances appear to give reasonable results \citep{Jennings18,Strader19}.

\section{Conclusions}
\label{sec:discussion}
We have presented the discovery of the suspected optical and X-ray counterparts to the unassociated \emph{Fermi} $\gamma$-ray source 4FGL J2333.1--5527, showing that it is likely a redback MSP binary harboring a massive neutron star with a heated companion in a nearly edge-on orbit. Our Keplerian fit to the radial velocity data suggests a primary mass well in excess of the canonical $\sim1.4\,M_{\odot}$, although tighter constraints on the binary mass ratio, from high-resolution optical spectroscopy to determine the projected rotational velocity of the secondary or future timing of the radio pulsar, are necessary to bolster this tentative conclusion.
%of $M_{NS} = 1.88 \pm 0.24$\,\Msun~making this potentially one of the most massive neutron stars found to date.
If the radio pulsar is found and timed, the likely edge-on geometry of this system would be ideal for measuring the relativistic Shapiro delay to precisely measure a massive neutron star
\citep[e.g.,][]{Demorest10, Antoniadis13, Cromartie19}, though unfortunately this is likely to be challenging due to radio eclipses.

The hard power-law X-ray spectrum in J2333 is similar to that of other recently discovered redback MSPs and is consistent with emission from an intrabinary shock. The inferred X-ray luminosity of $\sim 5 \times10^{31}$ erg s$^{\rm{-1}}$ is also in line with known redbacks in the pulsar state, even considering the uncertain distance \citep{Linares14, Strader19}. Future X-ray observations can better constrain the properties of the shock, especially given the reasonably well-understood inclination of the binary.

Although our optical and X-ray data provide compelling evidence that this object is a redback radio pulsar, the most significant advance towards fully understanding and characterizing the binary in 4FGL J2333.1--5527 would be the detection of radio pulsations at the position of the optical/X-ray source, followed by a precise pulsar timing solution. This would confirm our interpretation of the source and would place important constraints on the size of the orbit and mass ratio of the binary that are only weakly constrained at present by our modeling of the light and radial velocity curves. Such a detection would also enable a search for high energy pulsations modulated at the spin period of the binary as has been found in a number of other redback MSPs \citep[e.g., ][]{Johnson15, Smith17}. In the absence of radio pulsations, given the brightness of the source in $\gamma$-rays, a brute force $\gamma$-ray pulsation search could also be performed in order to definitively link the optical/X-ray object to the $\gamma$-ray source.

%Though the conclusions currently does mention searching for the radio pulsar and possible gamma-ray pulsations, I suggest that the authors reorder/rewrite the conclusions section to clarify what evidence supports which conclusions regarding the red back nature of the optical object, and its association with the gamma-ray source, and what kind of observations would be needed to provide direct evidence of the association. For example; the position and/or orbital parameters of a radio pulsar would link it to the optical object, while gamma-ray pulsations at the period and/or orbital parameters of the radio pulsar would link all three. In the absence of radio pulsations, a brute force gamma-ray pulsation search could link optical object to the gamma-ray source.

The unveiling of a new redback system among the brighter unassociated 4FGL sources suggests an exciting continuing discovery space for compact binaries with \emph{Fermi} in the months and years to come.

\section*{Acknowledgements}
We thank the anonymous referee for their thoughtful comments. We gratefully acknowledge support from NSF grant AST-1714825, NASA grant 80NSSC17K0507, and the Packard Foundation.

Based on observations obtained at the Southern Astrophysical Research (SOAR) telescope, which is a joint project of the Minist\'{e}rio da Ci\^{e}ncia, Tecnologia, Inova\c{c}\~{o}es e Comunica\c{c}\~{o}es (MCTIC) do Brasil, the U.S. National Optical Astronomy Observatory (NOAO), the University of North Carolina at Chapel Hill (UNC), and Michigan State University (MSU).

\bibliography{report}

\end{document}